\documentclass[namedreferences]{kluwer}
\usepackage[dvips]{epsfig}

\begin{document}
\begin{article}
\begin{opening}

\title{THE NORTH$-$SOUTH ASYMMETRY OF SOFT X-RAY FLARE INDEX DURING SOLAR
CYCLES 21, 22 AND 23}

\author{BHUWAN \surname{JOSHI} and ANITA JOSHI}

\institute{State Observatory, Naini Tal, 263 129, India (E-mail:
bhuwan@upso.ernet.in, anita@upso.ernet.in)}

\begin{ao}
Bhuwan Joshi\\
State Observatory,\\
Manora Peak, Naini Tal$-$263 129,\\
INDIA.\\
Phone: +91-05942-235583, 235136\\
Fax: +91-05942-235136\\
email: bhuwan@upso.ernet.in
\end{ao}

\begin{abstract}
In this paper the N$-$S asymmetry of the soft X-ray flare index 
($FI_{SXR}$) during the solar cycles 21, 22 and 23 has been analyzed.
The results show the existence of a real N$-$S asymmetry which is 
strengthened during solar minimum. The slope of the regression lines fitted to 
the daily values of asymmetry time series has been found to be negative
in all the three cycles. The yearly asymmetry curve can be fitted by 
a sinusoidal function with a period of eleven years. The power spectral analysis
of daily asymmetry time series reveals the significant periods of around 
28.26 days, 550.73 days and 3.72 years. 
\end{abstract}

\end{opening}

\section{Introduction}
It is well known that many types of solar phenomena exhibit some 
form of north-south (N$-$S) asymmetry. Bell (1962) found 
a long term asymmetry in the sunspot area data for cycle 8 through 
18. Roy (1977) studied the N$-$S distribution of major solar
flares from 1955 to 1974, covering cycles 19 and 20, and found 
an asymmetry in favour of northern hemisphere which is more pronounced
 during the minimum of solar cycle. Kno\v{s}ka (1985) studied the 
asymmetry of flare activity during the years 1937$-$1978 using 
the H$\alpha$ flare index ($FI_{H\alpha}$) introduced by Kleczek (1952, 1953)
and found no unique relationship between 
the asymmetry of flare activity and eleven year solar cycle. Vizoso
 and Ballester (1987) have studied the sudden disappearance (SD) 
of solar prominence during cycles 18$-$21 and found that 
the N$-$S asymmetry curve is not in phase with the solar cycle 
and peaks at or around the solar minimum and that the asymmetry 
changes its sign during the maximum, at the time of reversal of 
the Sun's general magnetic field. Verma (1993) reported a study of various
solar phenomena occurring in both northern and southern hemisphere
of the Sun for solar cycles 8$-$22. Joshi (1995); Li, Schimider, and 
Li (1998) and Ata\c{c} and \"{O}zg\"{u}\c{c} (1998) studied the N$-$S 
asymmetry during the solar cycle 22 with different manifestations
 of solar activity. In all these studies the asymmetry during 
this cycle was found to be in favour of southern hemisphere. Ata\c{c} and
 \"{O}zg\"{u}\c{c} (2001) studied the N$-$S asymmetry of $FI_{H\alpha}$
 during the 
rising phase of solar cycle 23 (1996$-$1999) and found that the 
activity prefers the northern hemisphere in general during the 
present cycle. Recently Li et al. (2002) have 
presented the dominant hemisphere of solar 
activity in each of solar cycle 12 to 22 by calculating the actual 
probability of the hemisphere distribution of sunspot groups, 
sunspot areas, the relative number of sunspots, flare index and number 
of SDs. 

Studies of periodicity in time series of N$-$S asymmetry data have also
been made by several authors. Vizoso and Ballester (1989) 
found a significant peak of around 3.1$-$3.2 years in the power 
spectrum performed with the values of flare number and flare 
index N$-$S asymmetry. Carbonell, Oliver, and Ballester (1993) and 
Oliver and Ballester (1994) studied the periodicity in the N$-$S 
asymmetry of sunspot area from 1874$-$1989 and 1874$-$1993 
respectively.  They obtained statistically significant period 
of 163.93 rotations (12.1 years). \"{O}zg\"{u}\c{c} and Ata\c{c} (1996) performed 
the power spectral analysis of N$-$S asymmetry time series of 
flare index values of solar cycle 22 and confirmed the 25.5 day 
fundamental period of the sun which was discovered by Bai and Sturrok (1991).
Taking all this into account, here we have made an attempt to 
investigate the N$-$S asymmetry of soft X-ray flare index ($FI_{SXR}$),
which is based on the continuous record of SXR flares observed
by GOES during solar cycles 21, 22 and 23. The daily values of $FI_{SXR}$
include the contribution of SXR flares of different classes
(B to X) and are suitable for a short and intermediate term solar 
activity analysis. In the paper we have also searched for the periodicity 
in the daily asymmetry time series of $FI_{SXR}$ for the period 1975$-$2003. 

\begin{figure}
\epsfig{file=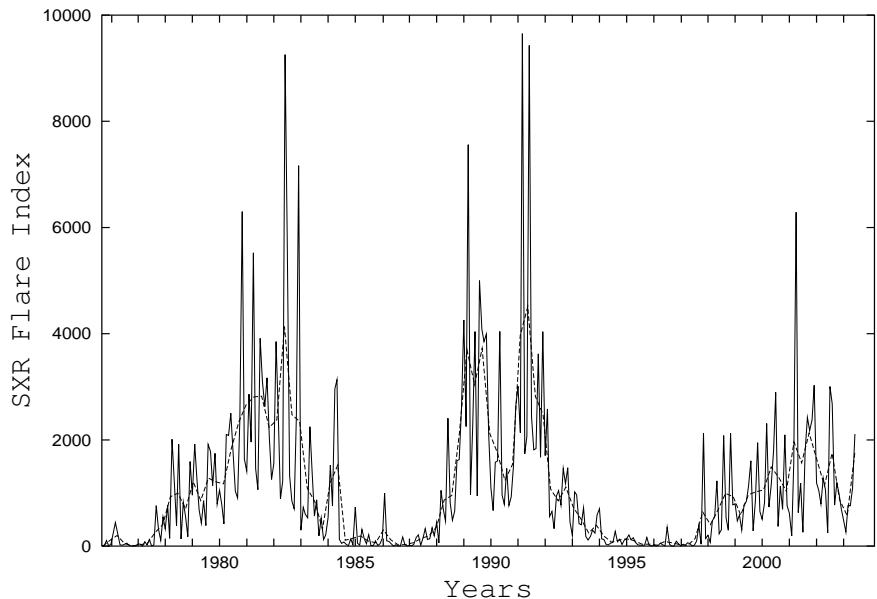,width=8.0cm, height=12.0cm,angle=270}
\vspace{0.5cm}
\caption{Plot of the monthly values of $FI_{SXR}$ for the period 1975$-$2003.
Dashed line shows the smooth spline curve.}
\end{figure}                                                                     

\section{The Soft X-ray Flare Index}
For the present analysis, first we have calculated daily $FI_{SXR}$
 by using the data of SXR flares observed by
GOES in 1$-$8 \AA\ wavelength band.
The data for the time span of 01 September 1975 to 30 June 2003 has been
downloaded from NGDC's anonymous ftp server: ftp://ftp.ngdc.no aa.gov/STP/
SOLAR$\_$DATA/SOLAR$\_$FLARES/XRAY$\_$FLARES. 
During this period the occurrence of 56,072 SXR flares are reported.

The $FI_{SXR}$ is introduced by Antalov\'a (1996) by 
weighing  the SXR flares of classes C, M and X as 1, 10 and 100
respectively (in units of $10^{-6}$ $Wm^{-2}$). In this manner, for example,
the SXR flare index for individual flares of class C7.3, M2.9 and X4.8
is 7.3, 29.0 and 480 respectively (See also Landi et al., 1998). 
Here, we have also included
the contribution of flares of class B in the calculation of daily
$FI_{SXR}$. Suppose $m_{B}$, $m_{C}$, $m_{M}$ and $m_{X}$ are the
digit multipliers for the flares of class B, C, M and X respectively,
then the expression to calculate daily $FI_{SXR}$ can be written
in the following way

\begin{figure}
\epsfig{file=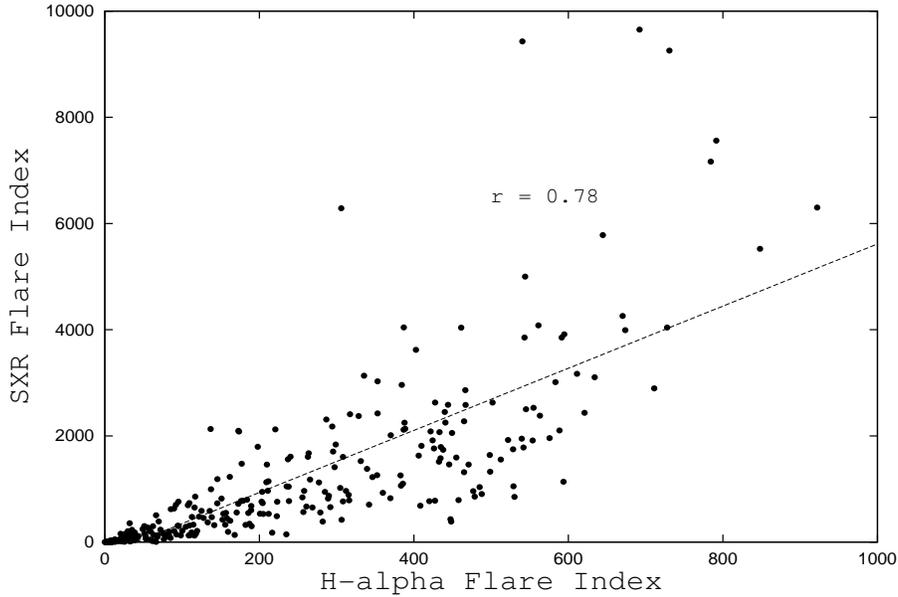,width=8.0cm, height=12.0cm,angle=270}
\vspace{0.5cm}
\caption{Scatter diagram showing the correlation between the monthly values of 
$FI_{SXR}$ and $FI_{H\alpha}$. Here r indicates the 
correlation coefficient.}
\end{figure}      

\begin{equation}
FI_{SXR} = 0.1 \times \sum_{i}^{} m_{B} + 1.0 \times \sum_{j}^{} m_{C} + 10.0 \times \sum_{k}^{} m_{M} + 100.0 \times \sum_{l}^{} m_{X}
\end{equation}
where $i$, $j$, $k$ and $l$ are the number of flares of class B, C, M and X per day respectively.  

Figure 1 shows the variation of monthly 
values of $FI_{SXR}$ for the period of 1975$-$2003 covering solar 
cycles 21, 22 and the rising and maximum phase of cycle 23. We have compared
$FI_{SXR}$ with $FI_{H\alpha}$. For this a correlation plot of these
activity indicators is given in Figure 2. The $FI_{H\alpha}$ values are 
available for general use in Kandilli observatory's and NGDC's anonymous 
ftp servers:
ftp://ftp.koeri.boun.edu.tr/pub/astr- onomy/flare$\_$index and ftp://ftp.ngdc.noaa.gov/STP/SOLAR$\_$DATA/ SOLAR$\_$FLARES/INDEX. 

\section{The Analysis of Asymmetry Time Series}

\begin{figure}
\epsfig{file=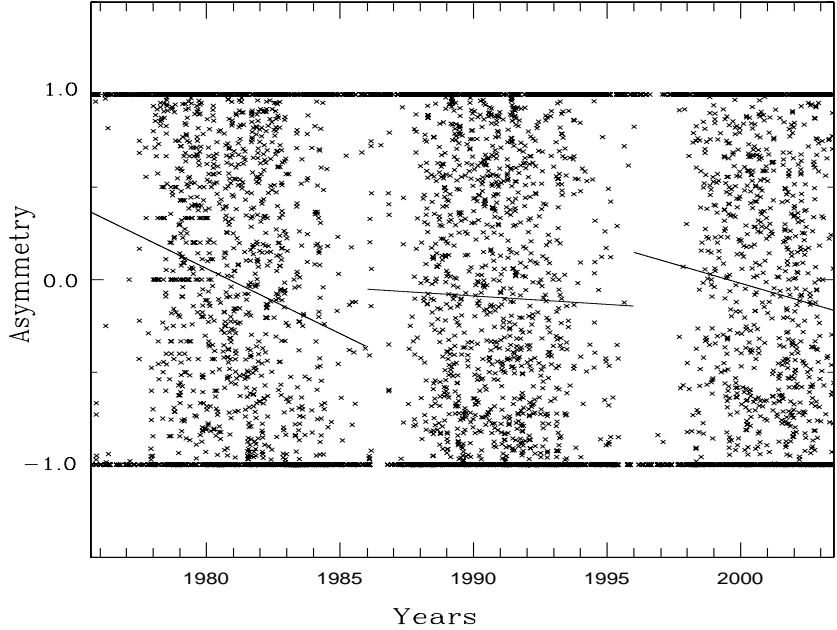, width=12cm, height=9.5cm}
\caption{ The plot of the daily values of asymmetry time series during 
solar cycles 21, 22 and 23. The straight line corresponds to the fit of first 
order polynomial.}
\end{figure}

The N$-$S asymmetry for the $FI_{SXR}$ is defined as
\begin{equation}
A_{SXR} = \frac{FI_{N} - FI_{S}}{FI_{N} + FI_{S}},
\end{equation}
where $FI_{N}$ and $FI_{S}$ stands for the daily $FI_{SXR}$ in the 
northern and southern solar hemisphere respectively. Thus, if 
$A_{SXR}$ \(>\)0, the activity in the northern hemisphere dominates
and if $A_{SXR}$ \(<\)0, the reverse is true. In the GOES flare
list the heliographic latitudes of all the flares are not known and SXR 
flares of class $\geq$ B 1.0 are not occurred daily.
Therefore such data were excluded from our study. Due to these 
reasons the above expression gives us an asymmetry time series 
composed of 6952 values. In Figure 3 the plot of daily asymmetry time
series is presented. To show the reality of the variations of asymmetry time 
series we have fitted a straight line to the daily values of $A_{SXR}$ for each
cycle separately.

\begin{figure}
\epsfig{file=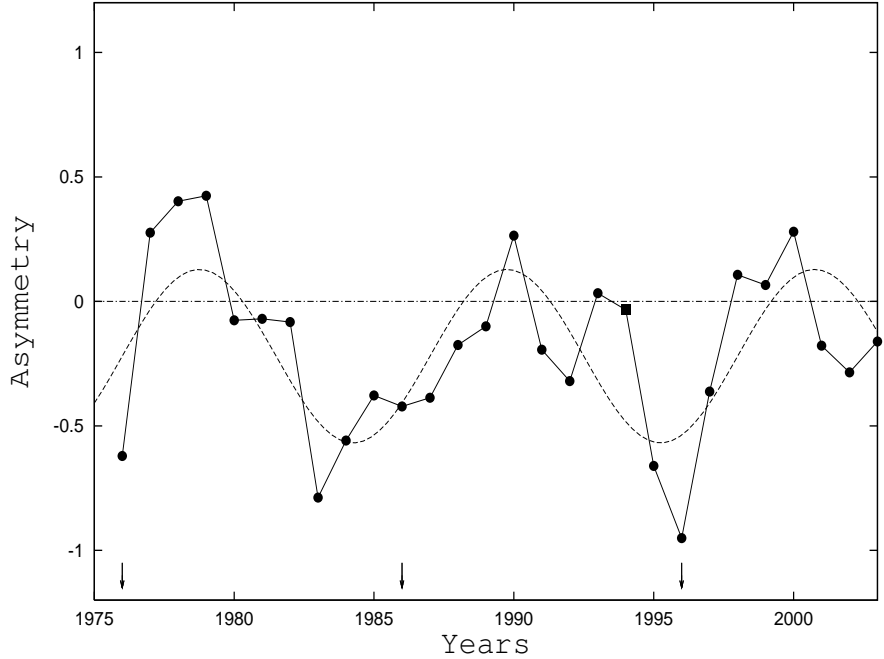, width=9cm, height=12cm,angle=270}
\vspace{0.5cm}
\caption{Yearly N$-$S asymmetry index of $FI_{SXR}$ during 1976$-$2003.
Highly significant asymmetry values with p$\geq$ 99.5 $\%$ are 
marked with circles, otherwise squares are drawn. Dashed line 
corresponds to the fitted sinusoidal curve with a period of eleven 
years. Downward arrows indicate solar activity minima.}
\end{figure}

To investigate to what extent the asymmetry 
is real we have followed the method of Letfus (1960) in which we can 
define the asymmetry of random distribution on the solar disk as
\begin{equation}
\Delta A_{SXR} = \pm \frac{1}{\sqrt{2(FI_{N} + FI_{S})}},
\end{equation}
which depends upon the values of $FI_{SXR}$ in the northern and
southern hemisphere respectively. To verify the reliability of 
calculated asymmetry values, $\chi$$^{2}$ test is applied with
\begin{equation}
\chi = \frac{2(FI_{N} - FI_{S})}{\sqrt{2(FI_{N} + FI_{S})}}=\frac{\sqrt{2} A_{SXR}}{\Delta A_{SXR}}.
\end{equation}
Thus for $A_{SXR}$ \(<\) $\Delta A_{SXR}$, $\Delta A_{SXR}$ $\leq$ $A_{SXR}$ \(<\) $2 \Delta A_{SXR}$ and $A_{SXR}$ $\geq$ $2 \Delta A_{SXR}$, the probability that N$-$S asymmetry exceeds the dispersion value
is p \(<\) 84 $\%$, 84 $\%$ $\leq$ p \(<\) 99.5 $\%$ and p $\geq$ 99.5
$\%$ respectively. Here the first, second and third limits imply
for the statistically insignificant, significant and highly significant
values respectively. Using the method of Letfus we have found 
that out of total 6952 cases, p $\geq$ 99.5 $\%$ in 4789 cases;
 84 $\%$ $\leq$ p \(<\) 99.5 $\%$ in 1237 cases and p \(<\) 84 $\%$
 in 926 cases, which means that, in general the N$-$S asymmetry
is a real phenomenon and not due to random fluctuations.
We have also studied the asymmetry of $FI_{SXR}$ 
year wise (see Figure 4). In this case we have 
found that out of 28 cases, the data is highly significant 
(p \(>\) 99.5 $\%$) in 27 cases. 
                                                                    
We are also interested to see any possible connection of the 
asymmetric behaviour of $FI_{SXR}$ with a period of eleven years. For
this, in Figure 4, a comparison between the original asymmetry curve
 and a fitted sinusoidal function is also presented. The
fitted sinusoidal function is of the form
\begin{equation}
y(x) = C + A \sin 2 \pi f (t-t_{0}),
\end{equation}
where $C$ is a constant term and $A$ is the amplitude. $t_{0}$ has been  
assigned a value 1976. The frequency $f$ has been given a fixed
value corresponding to eleven year period. Thus $f$ = 1/11 or 0.091 
years $^{-1}$. The values of $C$ and $A$ come out to be $-$0.22 and
0.35 respectively. We find that there is a good agreement between
original curve of asymmetry index and the fitted sinusoidal curve.
          
\section{Power Spectra of Asymmetry Time Series}
Using the Lomb-Scargle periodogram method (Lomb, 1976; Scargle, 
1982; Horne and Baliunas, 1986) for the daily asymmetry time series, which is plotted
in Figure 3, we have performed power spectra analysis. 
Figures 5a, 6a and 7a show the normalized power spectra
of asymmetry time series during 1975$-$2003
for the three frequency ranges of 116.0$-$463.0, 16.5$-$116.0 and 
7.7$-$16.5 nHz which corresponds to the period intervals of 100$-$25, 700$-$100
and 1500$-$700 days respectively. The $FI_{SXR}$ is not independent. Therefore 
the probability P of the power density at a given frequency being greater than K by chance is given by
\begin{equation}
P(z > K) = exp (-K/k),
\end{equation}
where the normalization factor $k$, which is due to event correlation,
can be determined empirically (Bai and Cliver, 1990).

\begin{figure}
\epsfig{file=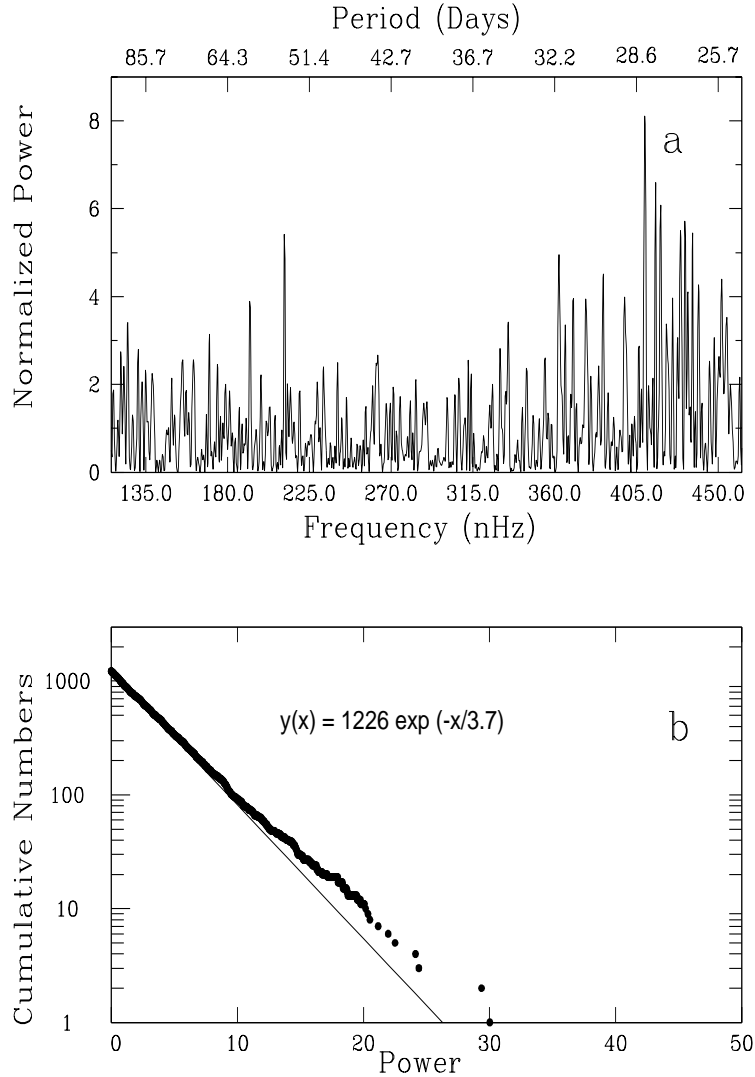, width=12.5cm, height=15cm}
\caption{(a) Normalized power spectrum of the daily asymmetry time 
series for the frequency interval of 116.0$-$463.0 nHz (100$-$25 days). (b)
Scargle power distribution corresponding to Figure 5a. The vertical
axis is the number of frequencies for which power exceeds x. The 
straight line is the fit to the points for lower values of power.}
\end{figure}

\begin{figure}
\epsfig{file=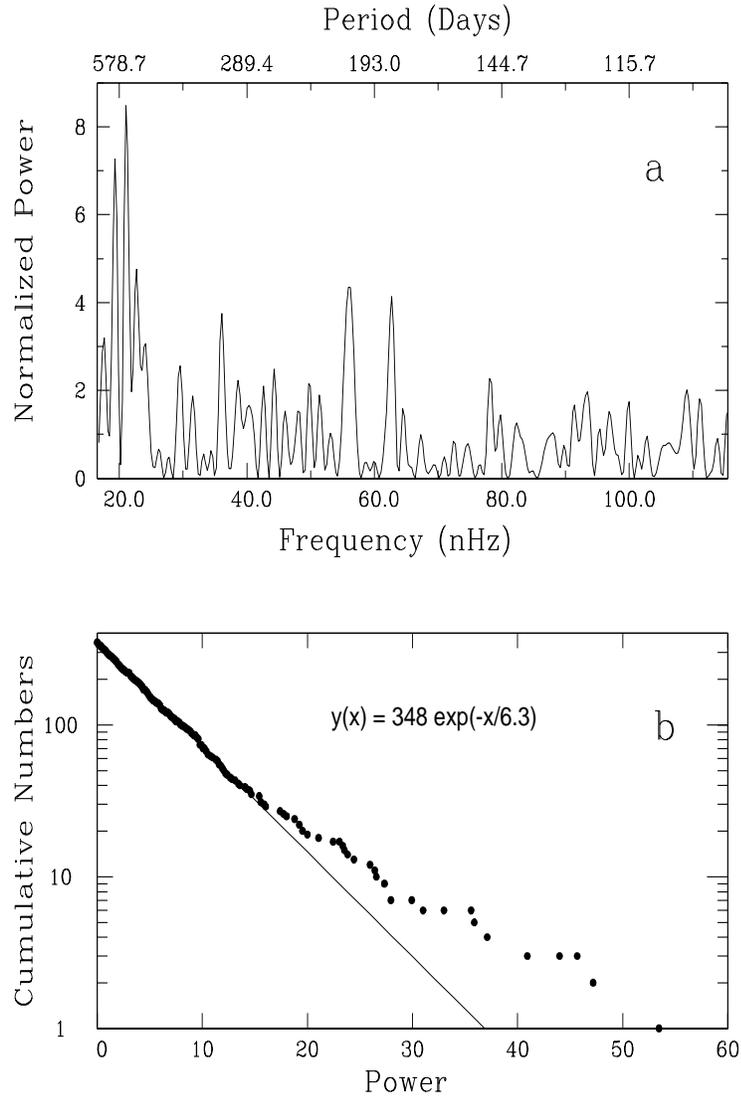, width=12.5cm, height=15cm}
\caption{Same as Figure 5, but for the frequency interval of 16.5$-$116.0
 nHz (700$-$100 days).} 
\end{figure}

\begin{figure}
\epsfig{file=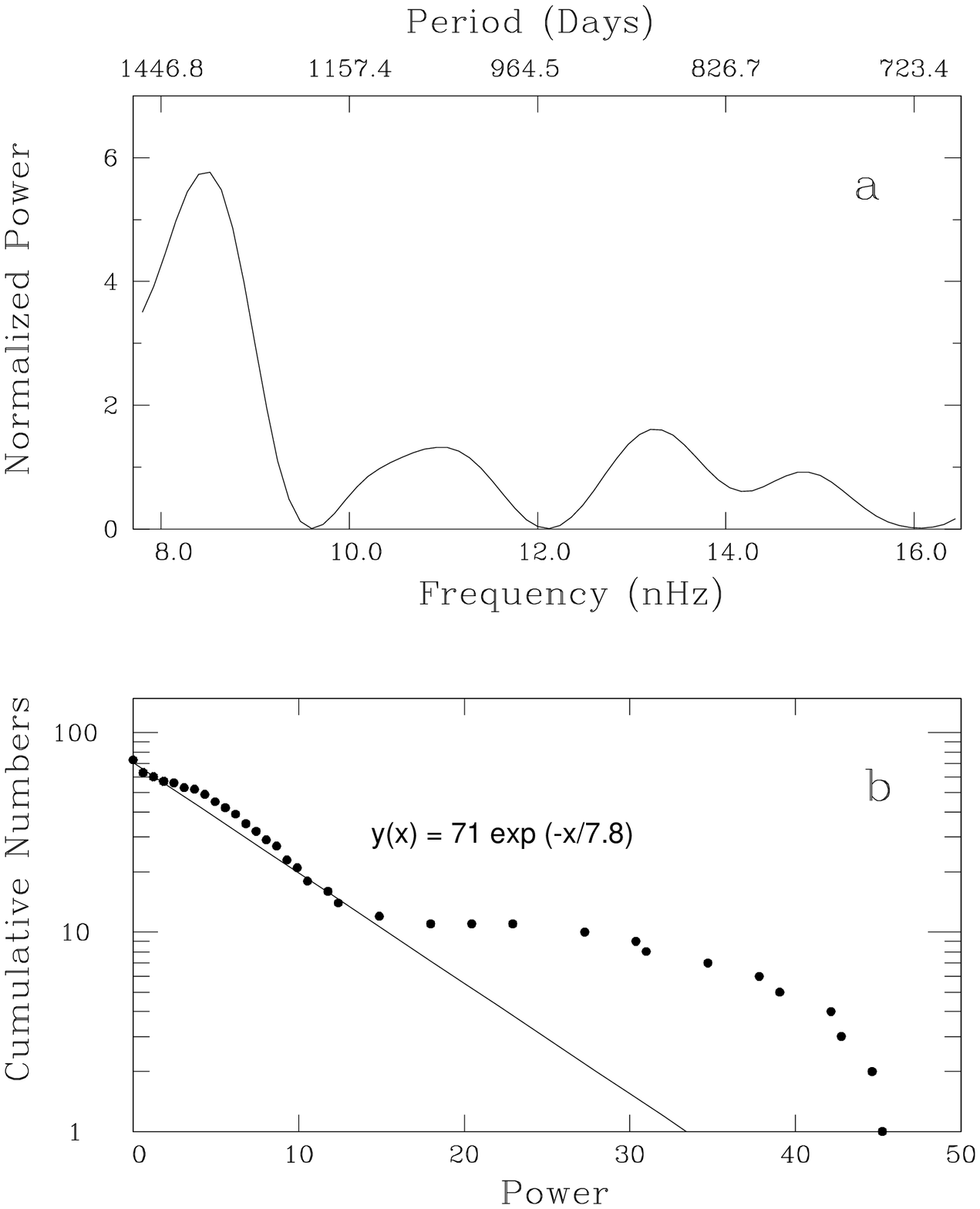, width=12.5cm, height=15cm}
\caption{Same as Figure 5, but for the frequency interval of 7.7$-$16.5
 nHz (1500$-$700 days).} 
\end{figure}

Figures 5b, 6b and 7b show the distribution of the power values
corresponding to the normalized power spectra shown in Figures
5a, 6a and 7a respectively. The vertical axis shows the cumulative 
number of frequencies
for which the power exceeds a certain value. For example, in the 
power spectrum for 116.0$-$463.0 nHz range shown in Figure 5a we have 1222
frequencies. For all these frequencies the power exceeds zero; thus, 
we have a point at (X = 0, Y = 1222). At only one frequency 
(409.5 nHz, which is equivalent to 28.26 days) the power was 30.02,
its maximum value. For  lower values of power, the distribution
can be well fitted by the equation y = 1226 exp(-x/3.7), as expected
from equation (6). Thus, we normalize the power spectrum by 
dividing the powers by 3.7 to obtain Figure 5a. For other cases we
use the same procedure for normalization.

\begin{figure}
\epsfig{file=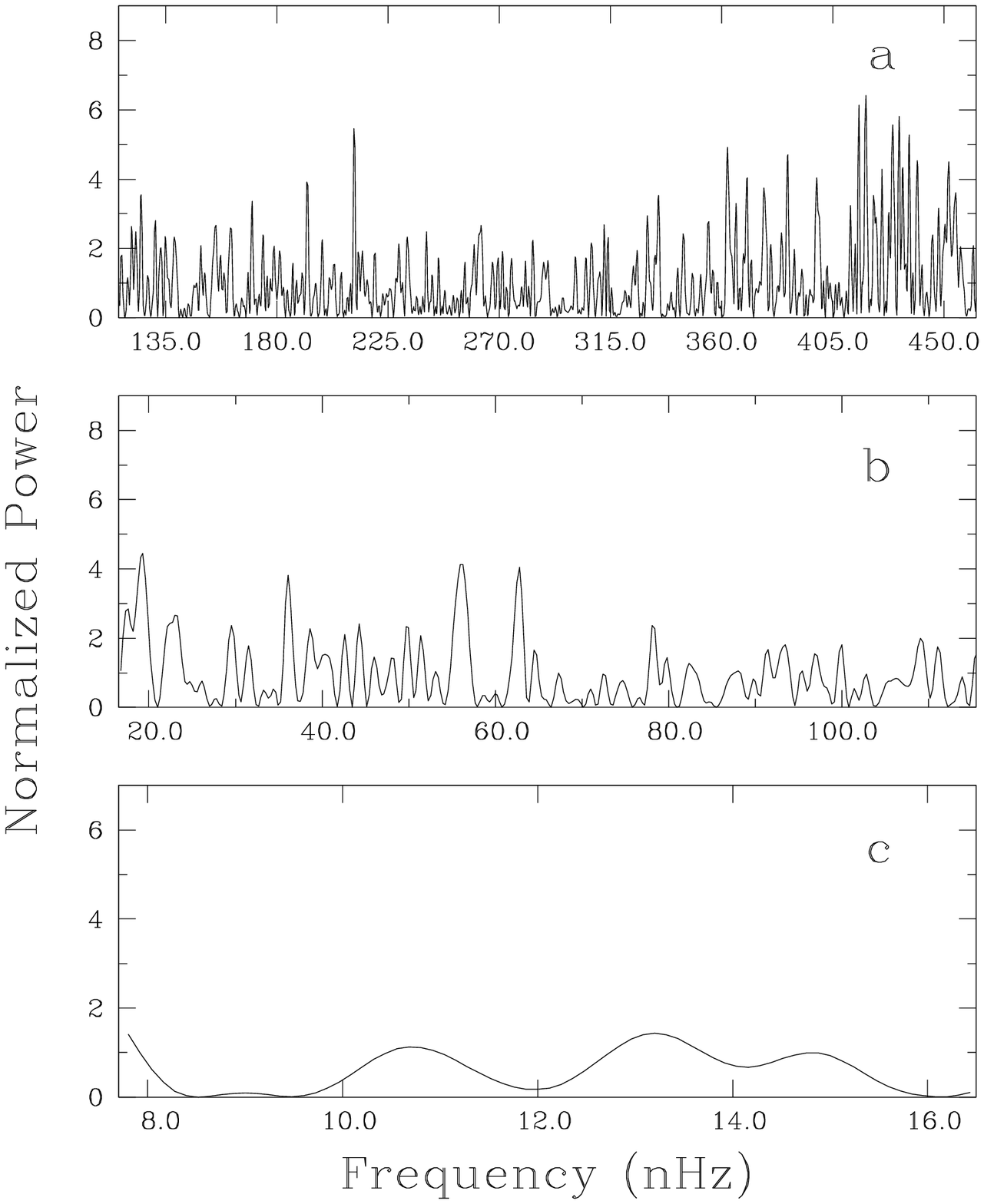, width=14.0cm, height=20cm}
\caption{ Periodogram of asymmetry time series, for the frequency intervals
of (a) 116.0$-$463.0 nHz (b) 16.5$-$116.0 nHz and (c) 7.7$-$16.5 nHz, after subtraction
of a sine curve with frequency 409.5 nHz (28.26 days), 21.02 nHz (550.73 days)
and 8.5 nHz (3.72 years) respectively.}
\end{figure}

Once the power spectrum has been normalized properly, we can use the
`False alarm probability' (FAP) formula to estimate the statistical
significance of a peak in the power spectrum. It is given by the 
expression
\begin{equation}
F = 1 - [1 - exp(-Z_{m})]^{N},
\end{equation}
where $Z_{m}$ is the height of the peak in the normalized power
spectrum and N is the number of independent frequencies.
Therefore, for example, in the case of
power spectrum shown in Figure 5a, if we substitute $Z_{m}$=8.13
and N=1222 (since we searched 1222 frequencies with 0.284 nHz
intervals) in equation (7) we get F=0.30, i.e., the probability
to obtain such a high peak at 409.5 nHz (28.26 days) by chance is about 30$\%$.
The same analysis has been applied to the power spectra in other two
frequency ranges. In the frequency interval of 16.5$-$116.0 nHz and 7.7$-$16.5 nHz
we get important peaks at 21.02 nHz (550.73 days) with a FAP =7$\%$ and 
8.5 nHz (3.72 years) with a FAP=20.4$\%$ respectively. 

To confirm the peaks of 28.26 days, 550.73 days and 3.72 years in 
Figures 5a, 6a and 7a respectively
are not due to aliasing, we have removed sine curves of these periods from 
the original time series. Then, once again, we have computed the periodogram
(see Figure 8).

\section{Discussions and Conclusions}
In this paper we have calculated the daily $FI_{SXR}$ for solar
cycles 21 to 23. The plot (Figure 1) of monthly values of $FI_{SXR}$
indicates a lower level of activity during solar cycle 23 compared
to previous two cycles. This has been pointed out earlier by 
\"{O}zg\"{u}\c{c} et al. (2002) in his study during the rising phase of solar
cycle 23. The scatter diagram (Figure 2) between $FI_{SXR}$ and
 $FI_{H\alpha}$ shows that there exists a good correlation between 
these two solar activity indices. Ata\c{c} and \"{O}zg\"{u}\c{c} (1989, 1998, 
2003) in their various papers have made extensive statistical analysis of 
$FI_{H\alpha}$, finding its good correlation with a number of solar indices 
that arise under different physical conditions. Therefore, a good 
correlation of $FI_{SXR}$ with $FI_{H\alpha}$ also makes the former an 
interesting parameter which is good enough to describe the solar activity. 

We have found that there exists a real N$-$S asymmetry in $FI_{SXR}$
which is not due to random fluctuations. The fitted straight line
to asymmetry time series (Figure 3) for the cycles 21 and 23 shows that the 
activity in the northern hemisphere is more important during the 
ascending branch of cycle whereas during the descending branch the 
activity becomes more important in the southern hemisphere. 
In the cycle 22 there was a southern dominance throughout. 
The slope of regression lines for all the three cycles is found to be
negative. Vizoso and Ballester (1990), in his study of N$-$S asymmetry 
of sunspot area during solar cycles 12$-$20 and sunspot number 
for cycle 21, found that the slope of regression lines fitted 
to the yearly values of N$-$S asymmetry change its sign each
four cycles suggesting a long term periodic behaviour in the 
N$-$S asymmetry of around eight cycles. They also found that the
slope of regression lines has changed from positive to negative
values after cycle 19. Ata\c{c} and \"{O}zg\"{u}\c{c} (1996) made similar 
conclusions with the N$-$S asymmetry of $FI_{H\alpha}$ during 
cycles 17$-$22. Our study confirms and extends these earlier results. 
Moreover, Ata\c{c} and \"{O}zg\"{u}\c{c} (1996) suggested that, if the period of 
eight cycles is real, the behaviour of asymmetry during cycles 
20$-$22 should be maintained in cycle 23, which has also been confirmed 
by the present study.

On the other hand, if we consider the plot of yearly N$-$S 
asymmetry index (Figure 4) the change in the predominance of activity
in the two hemisphere can be studied for every year starting from
1976. Initially during the solar cycle 21, the northern hemisphere
was more active but after 1980 it was shifted slightly towards south.
In the year 1983 the activity moved towards south strongly and
prevailed there during most of the solar cycle 22. In the solar 
cycle 23, we observe a strong peak in southern hemisphere during 
the minimum phase in the year 1996 and then the activity is moved
towards the north. In the years 1998, 1999 and 2000 the northern
hemisphere was dominated. These results are in agreement with the
work done by Temmer et al. (2001), who analyzed the N$-$S 
asymmetry of $H\alpha$ flares from 1975 to 1999. The preference
for northern hemisphere during the rising phase of cycle 23 is
also reported by Ata\c{c} and \"{O}zg\"{u}\c{c} (2001) with the data of 
$FI_{H\alpha}$. Here we also notice that, during cycle 23, the 
behaviour of asymmetry is changed in the year 2000 and it
has dominated towards southern hemisphere in the successive
years. The Figure also reveals that the fitted sinusoidal 
curve with a period of eleven years reflects the behaviour of 
asymmetry in a good manner. This suggests that N$-$S asymmetry of $FI_{SXR}$ has
a periodic behaviour shifted in phase with respect to solar cycle. 
The shape of the curve also indicates that the asymmetry has peaked
at or around the minimum of solar activity. 
This result is complementary to the study of N$-$S asymmetry
made by several authors (Swinson, Koyama and Saito, 1986; Vizoso
and Ballester, 1990; Joshi, 1995; Ata\c{c} and \"{O}zg\"{u}\c{c}, 1996).

Finally, we have studied periodicities in the daily asymmetry time 
series for three frequency ranges. In the power spectrum for the 
frequency range of 116.0$-$463.0 nHz (Figure 5), we get a significant 
peak at 409.5 nHz (28.26 days). It is close to the 27 days solar 
rotational period which is well known in different solar indices.
Antonucci et al. (1990) investigated the rotation of photospheric magnetic
field during solar cycle 21, and obtained a dominant period of 26.9 days
for the northern and 28.1 days for the southern hemisphere. A similar
outcome is obtained by Temmer et al. (2002) from the power spectral analysis
of daily sunspot numbers during 1975$-$2000 in which they derive a rigid
rotation with 27.0 days for the northern hemisphere, while the southern
hemisphere reveals a dominant period of 28.2 days. 
Pap, Tobiska, and Bouwer (1990) and Bouwer (1992) reported periods between
26$-$28 days from the power spectra of various solar indices.
 Joshi (1999) reported peaks at 26.5 days and 28.3 days in the 
power spectrum of solar radio flux at 10.7 cm. Lou et al. (2003) found a 
prominent peak of around 28 days in the daily averaged Ap index for geomagnetic
disturbances, which is
physically identified with solar rotation. 
The power spectrum for the 16.5$-$116.0
nHz range (Figure 6) shows a significant peak at 21.02 nHz 
(550.73 days). Ichimoto et al. (1985) reported that a peak in the 
power spectra appears near 510$-$540 days when northern and southern 
hemisphere flares are analyzed separately. Bai (1987) found an 
important peak at 552 days in the power spectra of major flares 
belonging to northern hemisphere. A significant peak of around 20 
rotations (about 540 days) is found by Akioka et al. (1987) in 
the power spectra for TA (Sum of maximum areas of sunspot groups 
per solar rotation) and MA (mean area of sunspot groups 
per solar rotations) during solar cycle 21. \"{O}zg\"{u}\c{c} and Ata\c{c} (1989)
reported a peak at 564 days in the $FI_{H\alpha}$ during cycle
20 and 21 considered together. Oliver, Carbonell, and Ballester (1992)
found the periodicities around 540 days in the cycles 12, 14 and 17 
in sunspot areas, while during cycles 18 and 19 it is present, with a
very high significance, in sunspot and Zurich sunspot number. They 
also found a peak at 528 days in the periodogram for sunspot areas 
during cycles 12 to 21. In the power spectra for the frequency range
of 7.7$-$16.5 nHz (Figure 7), we have a significant peak at 8.5 nHz 
(3.72 years). Rao (1973) studied periodicities in several indices
of solar activity and found that the index P (average area of all 
spot groups), $T_{0}$ (average spot group life time) and $f_{0}$ 
(number of all spot groups formed all over the Sun within a unit 
time) presented periodicities around 3.9, 3.5 and 3.48 years,
respectively. Vizoso and Ballester (1989) found a periodicity of 
around 3.1$-$3.2 years in the values of $FI_{H\alpha}$ and flare numbers 
N$-$S asymmetry. A periodicity of 3.27 years is reported
by Vizoso and Ballester (1990) in the N$-$S asymmetry of sunspot 
areas. In all the three power spectra presented by us, the 
subtraction of sine curve indicates that the peaks are probably 
real and not spurious or due to aliasing (Figure 8). We, therefore, 
see that the periodicities in the asymmetry time series of 
$FI_{SXR}$ is consistent with the earlier results.

\begin{acknowledgements}

We are thankful to Prof. Ram Sagar for valuable comments and suggestions.
One of the authors (BJ) wishes to thank Mr. Brijesh Kumar and Mr. J. C. Pandey 
for help in data analysis.
\end{acknowledgements}

\end{article}

\addcontentsline{toc}{section}{References}
\begin{thebibliography}{}  

\bibitem[\protect\citeauthoryear{Akioka et al.}{1987}]{Akioka87}
Akioka, M., Kubota, J., Suzuki, M., Ichimoto, K., and Tohmura, I.: 1987, {\it Solar Phys.} {\bf 112}, 313.                              

\bibitem[\protect\citeauthoryear{Antalova}{1996}]{Antalova96}
Antalov\'a, A.: 1996, {\it Contrib. Astron. Obs. Skalnat\'e Pleso.}
{\bf 26}, 98.








\bibitem[\protect\citeauthoryear{Antonucci}{1990}]{Antonucci90}
Antonucci, E., Hoeksema, J. T., and Scherrer, P. H. : 1990, 
{\it Astrophys. J.}
{\bf 360}, 296.

\bibitem[\protect\citeauthoryear{Atac et al.}{1996}]{Atac96}
Ata\c{c}, T. and \"{O}zg\"{u}\c{c}, A.: 1996, {\it Solar Phys.}
{\bf 166}, 201.

\bibitem[\protect\citeauthoryear{Atac et al.}{1998}]{Atac98}
Ata\c{c}, T. and \"{O}zg\"{u}\c{c}, A.: 1998, {\it Solar Phys.}
{\bf 180}, 397.

\bibitem[\protect\citeauthoryear{Atac et al.}{2001}]{Atac01}
Ata\c{c}, T. and \"{O}zg\"{u}\c{c}, A.: 2001, {\it Solar Phys.}
{\bf 198}, 399.

\bibitem[\protect\citeauthoryear{Bai}{1987}]{Bai87}
Bai, T.: 1987, {\it Astrophys. J.}
{\bf 318}, L85.

\bibitem[\protect\citeauthoryear{Bai}{1990}]{Bai90}
Bai, T. and Cliver, E. W..: 1990, {\it Astrophys. J.}
{\bf 363}, 299.

\bibitem[\protect\citeauthoryear{Bai et al.}{1991}]{Bai91}
Bai T. and Sturrock, P. A.: 1991, {\it Nature} {\bf 352}, 360.       

\bibitem[\protect\citeauthoryear{Bell}{1962}]{Bell 62}
Bell, B.: 1962, {\it Smithsonian Contr. Astrophys.}
{\bf 5}, 203.

\bibitem[\protect\citeauthoryear{Bouwer}{1992}]{Bouwer92}
Bouwer, S. D.: 1992, {\it Solar Phys.}
{\bf 142}, 365.

\bibitem[\protect\citeauthoryear{}{}]{}
Carbonell, M., Oliver, R., and Ballester, J. L.: 1993, {\it Astron. Astrophys.} {\bf 274}, 497.       

\bibitem[\protect\citeauthoryear{Horne}{1986}]{Horne86}
Horne, J. H. and Baliunas, S. L.: 1986, {\it Astrophys. J.}
{\bf 302}, 757.

\bibitem[\protect\citeauthoryear{Ichimoto}{1985}]{Ichimoto85}
Ichimoto, K., Kubota, J., Suzuki, M., Tohmura, I., and Kurokawa, H.: 1985, {\it Nature}
{\bf 316}, 422.

\bibitem[\protect\citeauthoryear{Joshi}{1995}]{Joshi95}
Joshi, A.: 1995, {\it Solar Phys.}
{\bf 157}, 315.

\bibitem[\protect\citeauthoryear{Joshi}{1999}]{Joshi99}
Joshi, A.: 1999, {\it Solar Phys.}
{\bf 185}, 397.

\bibitem[\protect\citeauthoryear{Atac et al.}{1998}]{Atac98}
Kleczek, J.: 1952, {\it Publ. Contr. Astron.}
{\bf No. 22}, Prague.

\bibitem[\protect\citeauthoryear{Atac et al.}{1998}]{Atac98}
Kleczek, J.: 1953, {\it Publ. Astrophys. Obs. Czech. Acad. Sci.}
{\bf No. 24}, Prague.

\bibitem[\protect\citeauthoryear{Atac et al.}{1998}]{Atac98}
Kno\v{s}ka, \v{S}.: 1985, {\it Contrib. Astron. Obs. Skalnat\'e Pleso}
{\bf 13}, 217.

\bibitem[\protect\citeauthoryear{Landi et al.}{1998}]{Landi98}
Landi, R., Moreno, G., Storini, M., and Antalov\'a, A.: 1998, {\it J. Geophys. Res.} {\bf 103}, No. A9, 20553.

\bibitem[\protect\citeauthoryear{Letfus}{191960}]{Letfus60}
Letfus, V.: 1960, {\it Bull. Astron. Inst. Czech.}
{\bf 11}, 31.

\bibitem[\protect\citeauthoryear{Atac et al.}{1998}]{Atac98}
Li, K. J., Schmieder, B., and Li, Q. S.: 1998, {\it Astron. Astrophys.}
{\bf 131}, 99.

\bibitem[\protect\citeauthoryear{Atac et al.}{1998}]{Atac98}
Li, K. J., Wang, J.X., Xiong, S. Y., Liang, H. F., Yun, H. S., and Gu, X. M.: 2002, {\it Astron. Astrophys.}
{\bf 383}, 648.

\bibitem[\protect\citeauthoryear{Lomb}{1976}]{Lomb76}
Lomb, N.: 1976, {\it Astrophys. Space Sci.}
{\bf 39}, 477.

\bibitem[\protect\citeauthoryear{Lou}{2003}]{Lomb76}
Lou, Y., Wang, Y., Fan, Z., Wang, S., and Wang, J.: 2003, {\it Monthly Notices 
Royal Astron. Soc.}
{\bf }, In press.

\bibitem[\protect\citeauthoryear{Oliver et al.}{1992}]{Oliver92}
Oliver, R., Carbonell, M., and Ballester, J.L.: 1992, {\it Solar Phys.}
{\bf 137}, 141.

\bibitem[\protect\citeauthoryear{Oliver et al.}{1994}]{Oliver94}
Oliver, R., Ballester, J.L.: 1994, {\it Solar Phys.}
{\bf 152}, 481.

\bibitem[\protect\citeauthoryear{Ozguc et al.}{1989}]{Ozguc89}
\"{O}zg\"{u}\c{c}, A. and Ata\c{c}, T.: 1989, {\it Solar Phys.}
{\bf 123}, 357.

\bibitem[\protect\citeauthoryear{Ozguc et al.}{1996}]{Ozguc96}
\"{O}zg\"{u}\c{c}, A. and Ata\c{c}, T.: 1996, {\it Solar Phys.}
{\bf 163}, 183.

\bibitem[\protect\citeauthoryear{Ozguc et al.}{2002}]{Ozguc02}
\"{O}zg\"{u}\c{c}, A., Ata\c{c}, T., and Ryb\'{a}k, J.: 2002, {\it J. Geophys. Res.}
{\bf 107}, No. A7, 10.1029/2001JA009080.

\bibitem[\protect\citeauthoryear{Ozguc et al.}{2003}]{Ozguc03}
\"{O}zg\"{u}\c{c}, A., Ata\c{c}, T., and Ryb\'{a}k, J.: 2003, {\it Solar Phys.}
{\bf 214}, 375.

\bibitem[\protect\citeauthoryear{Pap et al.}{1990}]{Pap90}
Pap, J., Tobiska, W.K., and Bouwer, S. D.: 1990, {\it Solar Phys.}
{\bf 129}, 165.

\bibitem[\protect\citeauthoryear{Rao}{1973}]{Rao73}
Rao, R.: 1973, {\it Solar Phys.}
{\bf 29}, 47.

\bibitem[\protect\citeauthoryear{Roy}{1977}]{Roy77}
Roy, J. R.: 1977, {\it Solar Phys.}
{\bf 52}, 53.

\bibitem[\protect\citeauthoryear{Scargle}{1982}]{Scargle82}
Scargle. J. D.: 1982, {\it Astrophys. J.}
{\bf 263}, 835.

\bibitem[\protect\citeauthoryear{Swinson}{1986}]{Swinson86}
Swinson, D. B., Koyama, H, and Saito, T.: 1986, {\it Solar Phys.}
{\bf 106}, 35.

\bibitem[\protect\citeauthoryear{Temmer et al.}{2001}]{Temmer01}
Temmer, M., Veronig, A., Hanslmeier, A., Otruba, W., and Messerotti, 
M..: 2001, {\it Astron. Astrophys.}
{\bf 375}, 1049.

\bibitem[\protect\citeauthoryear{Temmer et al.}{2002}]{Temmer02}
Temmer, M., Veronig, A., and Hanslmeier, A..: 2002, {\it Astron. 
Astrophys.}
{\bf 390}, 707.

\bibitem[\protect\citeauthoryear{Verma.}{1993}]{Verma93}
Verma, V. K.: 1993, {\it Astrophys. J.}
{\bf 403}, 797.

\bibitem[\protect\citeauthoryear{Vizoso et al.}{1987}]{Vizoso87}
Vizoso, G. and Ballester, J. L.: 1987 {\it Solar Phys.}
{\bf 112}, 317.

\bibitem[\protect\citeauthoryear{Vizoso et al.}{1989}]{Vizoso89}
Vizoso, G. and Ballester, J. L.: 1989 {\it Solar Phys.}
{\bf 119}, 411.

\bibitem[\protect\citeauthoryear{Vizoso et al.}{1990}]{Vizoso90}
Vizoso, G. and Ballester, J. L.: 1990 {\it Astron. Astrophys.}
{\bf 229}, 540.

\end{thebibliography}
\end{document}